\documentclass[aps,prd,nofootinbib,onecolumn,notitlepage]{revtex4-1}
\usepackage[utf8]{inputenc}
\usepackage{amsmath}
\usepackage{amsfonts}
\usepackage{amssymb}
\usepackage{color}
\usepackage[unicode=true,pdfusetitle,
 bookmarks=true,bookmarksnumbered=false,bookmarksopen=false,
 breaklinks=false,backref=false,colorlinks=false]
 {hyperref}

\renewcommand\[{\begin{equation}}
\renewcommand\]{\end{equation}}
\newcommand{\al}{\alpha}
\newcommand{\bt}{\beta}
\newcommand{\ga}{\gamma}

\newcommand{\n}{\nabla}
\newcommand{\ba}{\begin{eqnarray}}
\newcommand{\ea}{\end{eqnarray}}

\begin{document}
\title{Schwarzschild $1/r$-singularity is not permissible in ghost free quadratic curvature infinite derivative gravity }
\author{Alexey S. Koshelev$^{1,2,3,4}$,~Jo\~ao Marto$^{1,2}$,~Anupam Mazumdar$^{5,6}$}
\affiliation{$^{1}$Departamento de F\'isica, Universidade da Beira Interior, Rua Marquês D'Ávila e Bolama, 6201-001 Covilhã, Portugal.}
\affiliation{$^{2}$Centro de Matemática e Aplicações da Universidade da Beira Interior
(CMA-UBI), Rua Marquês D'Ávila e Bolama, 6201-001 Covilhã, Portugal.}
\affiliation{$^{3}$Theoretische Natuurkunde, Vrije Universiteit Brussel.} 
\affiliation{$^{4}$The International Solvay Institutes, Pleinlaan 2, B-1050, Brussels, Belgium.}
\affiliation{$^{5}$Van Swinderen Institute, University of Groningen, 9747 AG, Groningen, The Netherlands}
\affiliation{$^{6}$Kapteyn Astronomical Institute, University of Groningen, 9700 AV Groningen, The Netherlands.}
\date{\today}
\begin{abstract}
In this paper we will study the complete equations of motion for a ghost free quadratic curvature infinite derivative gravity. We will
argue that within the scale of non-locality, Schwarzschild-type singular metric solution is not {\it permissible}. Therefore, 
Schwarzschild-type vacuum solution which is a prediction in Einstein-Hilbert gravity 
may {\it not} persist within the region of non-locality. We will also show that just quadratic curvature gravity, without infinite derivatives,
always allows Schwarzschild-type singular metric solution.
\end{abstract}
 \vskip -1.0 truecm

\maketitle

\section{Introduction}

\label{Introduction }

Arguably Einstein's theory of general relativity is one of the most successful description of spacetime which has seen
numerous confirmations of observational tests at different length scales, predominantly in the infrared  (IR) (far away from the source 
and at late time scales)~\cite{-C.-M.}, including the fascinating detection of gravitational waves~\cite{-B.-P.}. 
In spite of this success, at short distances and at small time scales, i.e. in the ultraviolet (UV), 
the Einstein-Hilbert action leads to well-known singular solutions, in terms of blackhole solutions in the vacuum, and the cosmological 
singularity in a time dependent background~\cite{Hawking:1973uf}. The nature of latter singularity is indeed very different from the former, which 
brings uncertainty in the cosmological models at the level of initial conditions for inflation and the big bang cosmology.
In reality, one would expect that nature would avoid any kind of classical singularities, whether they are covered by an event horizon or 
they are naked - a stronger version of the {\it cosmic censorship} hypothesis~\cite{Penrose,Penrose:1964wq}.
In this respect, it can be argued that the singularities present in the Einstein-Hilbert action are mere artefacts of the action, and there must be a way
to  ameliorate the singularities in nature. Indeed, removing the singularities is one of the foremost fundamental questions of gravitational physics.

Recently, Biswas, Gerwick, Koivisto and Mazumdar (BGKM) have shown that the  quadratic curvature infinite derivative theory of gravity in $4$ 
spacetime dimensions can be made {\it ghost free} and avoid both cosmological and blackhole singularities at the linearised level around the 
Minkowski background~\cite{Biswas:2011ar}~\footnote{See previous to this work other relevant references~\cite{Tseytlin:1995uq,Siegel:2003vt,Biswas:2005qr}, 
where the authors have argued absence of singularity in infinite derivative gravity motivated from the string theory, however, the full quadratic curvature action including the Weyl term with two gravitational metric potentials were first presented in~\cite{Biswas:2011ar}.},
while the cosmological singularity can be resolved even at the full non-linear level~\cite{Biswas:2005qr,Biswas:2006bs,Koivisto,Koshelev:2012qn,Conroy:2014dja}. At the linear level (around 
asymptotically Minkowski background), resolution of blackhole singularities has been studied both in the context of static~\cite{Biswas:2011ar,Biswas:2013cha,Frolov-1,Edholm:2016hbt,Frolov,Koshelev:2017bxd,Buoninfante:2018xiw}, and in a 
rotating case~\cite{Cornell:2017irh} by various groups. Furthermore, lack of formation of singularity at the linear level has also been studied in a dynamical context 
by Frolov and his collaborators~\cite{Frolov:2015bia,Frolov:2015usa}.

In Ref.~\cite{Biswas:2011ar}, the authors have shown that for a {\it ghost free} quadratic curvature gravitational {\it form factors}, at short distances
the gravitational metric-potential tends to be a constant, while at large distances from the source the metric potential takes the usual form of $1/r$ behavior in the IR.
Furthermore, the gravitational force quadratically vanishes towards the center in the UV. Such a system behaves very much like a compact object, but by construction there is 
no curvature singularity, nor there is an event horizon. The gravitational entropy calculated by the Wald's formalism~\cite{Wald} leads to the {\it Area}-law~\cite{Conroy:2015wfa}. Since, all the interactions are {\it purely} derivative in nature, the gravitational {\it form factors} give rise to non-local interactions for such a spacetime~\cite{Tomboulis,Tomboulis:2015,Modesto,Talaganis:2014ida,Biswas:2014yia}. The non-locality is indeed {\it confined} within the scale $M_{s}$, which has very a interesting behavior. 


The aim of this short note is to show that the full non-linear metric solution of the BGKM gravity will not permit $1/r$-type metric potential, i.e. Schwarzschild type solution, for the static background. Note that what is relevant for us is indeed $1/r$ part of the metric potential, be it in isotropic coordinates or Schwarzschild's coordinates. Near the vicinity of singularity, at $r=0$, what dominates is indeed the $1/r$ part of the metric potential in Einstein's theory of gravity. Also, such a singular solution exists in quadratic curvature gravity as well~\cite{-K.-S.}, see for instance~\cite{Lu:2015cqa}, therefore it is a pertinent question to ask whether $1/r$-kind of metric potential would survive infinite derivative theory of gravity or not?


\section{The infinite covariant derivative action}

The most general quadratic curvature action (parity invariant and free from torsion) has been derived around constant curvature backgrounds in Refs.~\cite{Biswas:2011ar,Biswas:2013cha,Biswas:2016etb}, given by~
\footnote{The original action was first written in terms of the Riemann, but it is useful to write the action in terms of the Weyl term which is related to the Riemann as: \[
\label{weyl}
W_{\;\alpha\nu\beta}^{\mu}=R_{\;\alpha\nu\beta}^{\mu}-\frac{1}{2}(\delta_{\nu}^{
\mu}R_{\alpha\beta}-\delta_{\beta}^{\mu}R_{\alpha\nu}+R_{\nu}^{\mu}g_{
\alpha\beta}-R_{\beta}^{\mu}g_{\alpha\nu})+\frac{R}{6}(\delta_{\nu}^{\mu}g_{
\alpha\beta}-\delta_{\beta}^{\mu}g_{\alpha\nu})
\]
}
\begin{equation}\label{action}
S=\frac{1}{ {16\pi G} }\int d^{4}x\sqrt{-g}\left(R+\alpha_c\left[R{\cal
F}_{1}(\Box_s)R+R^{\mu\nu}{\cal
F}_{2}(\Box_s)R_{\mu\nu}+W^{\mu\nu\lambda\sigma}{\cal
F}_{3}(\Box_s)W_{\mu\nu\lambda\sigma}\right]\right)\,,
\end{equation}
where $G=1/M_p^2$ is the Newton's gravitational constant, $\alpha_c \sim 1/M_s^2$ is a dimensionful coupling, $\Box_s\equiv \Box/M_s^2$,  where $M_s$ 
signifies the scale of non-locality at which new gravitational interaction becomes important. In the limit $M_s\rightarrow \infty$, the action reduces to the 
Einstein-Hilbert term. The d'Alembertian term is $\Box=g^{\mu\nu}\nabla_{\mu}\nabla_{\nu}$, where $\mu\,,\nu=0,1,2,3$, and we work with a metric convention which is mostly positive $(-,+,+,+)$. The ${\cal F}_{i}$'s are three gravitational {\it form-factors}, 
{
\begin{equation}
{\cal F}_{i}(\Box_s) =\sum_{n} f_{i, n}\Box_s^{n}\,,
\end{equation}
}
reminiscence to any massless theory possessing {\it only} derivative interactions. In this theory, the graviton remains massless with transverse and traceless 
degrees of freedom. However, the gravitational interactions are non-local due to the presence of the form factors ${\cal F}_{i}$'s, see~\cite{Biswas:2011ar}. These form factors contain infinite covariant derivatives, which shows that the interaction vertex in this class of theory becomes non-local. In fact, the gravitational interaction in this class of theory leads to smearing out the point source by modifying the gravitational potential, as shown in \cite{Biswas:2011ar}. The non-local gravitational interactions are also helpful to ameliorate the quantum aspects of the theory, which is believed to be UV finite~\cite{Tomboulis,Tomboulis:2015,Modesto,Talaganis:2014ida}. The 
scale of non-locality is governed by $M_s^{-1}$. 

Around the Minkowski background the three {\it form factors} obey a constraint equation, in order to maintain {\it only} the transverse-and traceless graviton degrees of freedom, i.e.
the perturbative tree-level unitarity~\cite{Biswas:2011ar,Biswas:2013cha}
~\footnote{In order to make sure that the full action Eq.~(\ref{action}) contains the same original dynamical degrees of freedom as that of the massless graviton in $4$ dimensions.
This is to make sure that the action is {\it ghost free}, there are no other dynamical degrees of freedom in spite of the fact that there are infinite derivatives. The graviton propagator
for the above action gives rise to $$\Pi(k^2) =\frac{1}{a(k^2)}\Pi(k^2)_{GR} = \frac{1}{a(k^2)}\left[\frac{P^{(2)}}{k^2}-\frac{P^{(0)}}{2k^2}\right], $$ where  $P^{(2)}$ and $P^{(0)}$ are spin-2 and 0 projection operators, and $a(k^2)=e^{\gamma(k^2/M_s^2)}$, is exponential of an {\it entire function} - $\gamma$, which does not contain any 
poles in the complex plane, therefore no new degrees of freedom other than the transverse and traceless graviton, see for details~\cite{Biswas:2011ar,Biswas:2013kla}.  The gravitational form factors ${\cal F}_i(\Box_s)$ cannot be determined simultaneously in terms of  $a(\Box_s)$,  if we switch one of the  ${\cal F}_i=0$, then we can express the other form factors in terms of $a(\Box_s)$, for instance for ${\cal F}_2=0$ yields, ${\cal F}_1=-[(a(\Box_s)-1)/12\Box_s]$ and ${\cal F}_3=[(a(\Box_s)-1)/4\Box_s]$, see~\cite{Biswas:2013cha}.}
\begin{equation}
6{\cal F}_1(\Box_s)+3 {\cal F}_2(\Box_s)+ 2{\cal F}_3(\Box_s)=0\,.
\end{equation}
Let us first discuss very briefly the linear properties of this theory around an asymptotically Minkowski background before addressing the non-linear equations of motion. The linear solutions are indeed insightful and provides a lot of understanding of the solutions within BGKM gravity. Even though, we will not discuss explicitly non-linear solution, but any non-linear solution should have a limit in the linear regime. Note that the mass of the source is the relevant parameter, which plays a crucial role in determining linear and non-linear solutions.  In Refs.~\cite{Biswas:2011ar,Edholm:2016hbt,Frolov-1}, it was shown that for $a(\Box_s)= e^{\gamma(\Box_s)}$, where $\gamma$ is an {\it entire function}, the central singularity is avoided, while recovering the correct $1/r$ dependence in the metric potential in the IR. For a specific choice of $a(\Box_s)=e^{\Box_s}$,  and assuming the Dirac-delta mass distribution, $m\delta^3(r)$ at the center, the gravitational metric potential, i.e. the Newtonian potential remains linear, as long as:
\begin{equation}
mM_s \leq M_p^2\,,
\end{equation}
with the gravitational metric potential in static and  isotropic coordinates is given by~\cite{Biswas:2011ar}:
\begin{equation}\label{erf}
\phi(r)=-\frac{Gm}{r}{\rm Erf}\left(\frac{r M_s}{2}\right)\,,
\end{equation}
approaches to be constant with a magnitude less than 1 for $r< 2/M_s$. Since the error function goes linearly in $r$ for 
$r< 2/M_s$, the metric potential becomes finite in this ultraviolet region. For $ r > 2/M_s$, the metric potential follows as $\sim Gm/r$, in the infrared region. However, in our case the typical scale of non-locality is actually larger than the Schwarzschild's radius as shown in~\cite{Koshelev:2017bxd,Buoninfante:2018xiw}
\begin{equation}
r_{NL} \sim \frac{2}{M_s} \geq r_{sch} =\frac{2m}{M_p^2}\,,
\end{equation}
thus avoiding the event horizon as well
~\footnote{This could potentially resolve the 
information-loss paradox, since there is no event horizon and the graviton interactions for $r_{NL}\sim 2/M_s$ becomes nonlocal, therefore for interacting gravitons
the spacetime ceases to hold any meaning in the Minkowski sense.}. Now, for the rest of the discussion, let us focus on the full non-linear equations for the above action Eq.~(\ref{action}).

\section{Towards impossibility of the Schwarzschild metric solution}

The complete equations of motion have been derived from action Eq.~(\ref{action}), and they are given by \cite{Biswas:2013cha},
\begin{align}
P^{\alpha\beta}=& -\frac{G^{\alpha\beta}}{{ 8\pi G}} +\frac{\alpha_c}{{ 8\pi G}} \biggl( 4G^{\alpha\beta}{\cal
F}_{1}(\Box_s)R+g^{\alpha\beta}R{\cal
F}_1(\Box_s)R-4\left(\triangledown^{\alpha}\nabla^{\beta}-g^{\alpha\beta}
\Box_s\right){\cal F}_{1}(\Box_s)R
\nonumber\\&
-2\Omega_{1}^{\alpha\beta}+g^{\alpha\beta}(\Omega_{1\sigma}^{\;\sigma}+\bar{
\Omega}_{1}) +4R_{\mu}^{\alpha}{\cal F}_2(\Box_s)R^{\mu\beta}
\nonumber\\&
-g^{\alpha\beta}R_{\nu}^{\mu}{\cal
F}_{2}(\Box_s)R_{\mu}^{\nu}-4\triangledown_{\mu}\triangledown^{\beta}({\cal
F}_{2}(\Box_s)R^{\mu\alpha})
+2\Box_s({\cal
F}_{2}(\Box_s)R^{\alpha\beta})
\nonumber\\&
+2g^{\alpha\beta}\triangledown_{\mu}\triangledown_{
\nu}({\cal F}_{2}(\Box_s)R^{\mu\nu})
-2\Omega_{2}^{\alpha\beta}+g^{\alpha\beta}(\Omega_{2\sigma}^{\;\sigma}+\bar{
\Omega}_{2}) -4\Delta_{2}^{\alpha\beta}
\nonumber\\&
-g^{\alpha\beta}W^{\mu\nu\lambda\sigma}{\cal
F}_{3}(\Box_s)W_{\mu\nu\lambda\sigma}+4W_{\;\mu\nu\sigma}^{\alpha}{\cal {\cal
F}}_{3}(\Box_s)W^{\beta\mu\nu\sigma}
\nonumber\\&
-4(R_{\mu\nu}+2\triangledown_{\mu}
\triangledown_{\nu})({\cal {\cal F}}_{3}(\Box_s)W^{\beta\mu\nu\alpha})
-2\Omega_{3}^{\alpha\beta}+g^{\alpha\beta}(\Omega_{3\gamma}^{\;\gamma}+\bar{
\Omega}_{3}) -8\Delta_{3}^{\alpha\beta} \biggr)
\nonumber\\
=& -T^{\al\bt}\,,
\label{EOM}
\end{align}
where $T^{\al\bt}$ is the stress energy tensor for the matter components, and we have defined the following symmetric tensors, for the detailed derivation, see~\cite{Biswas:2013cha}:
\begin{align}\label{details}
\Omega_{1}^{\alpha\beta}= & \sum_{n=1}^{\infty}f_{1_{n}}\sum_{l=0}^{n-1}\nabla^{
\alpha}R^{(l)}\nabla^{\beta}R^{(n-l-1)},\quad\bar{\Omega}_{1}=\sum_{n=1}^{\infty
}f_{1_{n}}\sum_{l=0}^{n-1}R^{(l)}R^{(n-l)},
\\
\Omega_{2}^{\alpha\beta}= & \sum_{n=1}^{\infty}f_{2_{n}}\sum_{l=0}^{n-1}R_{\nu}^{
\mu;\alpha(l)}R_{\mu}^{\nu;\beta(n-l-1)},\quad\bar{\Omega}_{2}=\sum_{n=1}^{
\infty}f_{2_{n}}\sum_{l=0}^{n-1}R_{\nu}^{\mu(l)}R_{\mu}^{\nu(n-l)}\,,
\\
\Delta_{2}^{\alpha\beta}= & \sum_{n=1}^{\infty}f_{2_{n}}\sum_{l=0}^{n-1}
[R_{
\sigma}^{\nu(l)}R^{(\beta\sigma;\alpha)(n-l-1)}-R_{\;\sigma}^{\nu;\alpha(l)
}R^{
\beta\sigma(n-l-1)}]_{;\nu}\,,
\\
\Omega_{3}^{\alpha\beta}= & \sum_{n=1}^{\infty}f_{3_{n}}\sum_{l=0}^{n-1}W_{
\: \: \nu\lambda\sigma}^{\mu;\alpha(l)}W_{\mu}^{\;\nu\lambda\sigma;\beta(n-l-1)},
\quad\bar{\Omega}_{3}=\sum_{n=1}^{\infty}f_{3_{n}}\sum_{l=0}^{n-1}W_{
\: \: \nu\lambda\sigma}^{\mu(l)}W_{\mu}^{\;\nu\lambda\sigma(n-l)}\,,
\\
\Delta_{3}^{\alpha\beta}= & \sum_{n=1}^{\infty}f_{3_{n}}\sum_{l=0}^{n-1}
[W_{\quad\sigma\mu}^{\lambda\nu(l)}W_{\lambda}^{\;\beta\sigma\mu;\alpha(n-l-1)}
-W_{\quad\sigma\mu}^{\lambda\nu\;\;;\alpha(l)}W_{\lambda}^{
\: \beta\sigma\mu(n-l-1)}
]_{;\nu}\,.\label{details-1}
\end{align}
The notation $\mathcal{R}^{(l)}\equiv\Box^l\mathcal{R}$ has been used for the curvature tensors and their covariant derivatives.
The trace equation is much more simple, and just for the purpose of illustration, we write it below~\cite{Biswas:2013cha}:
\begin{align}
P = &\frac{R}{{ 8\pi G}}+ \frac{\alpha_c}{{8\pi G}} \biggl( 12\Box_s{\cal F}_{1}(\Box_s)R+2\Box_s({\cal
F}_{2}(\Box_s)R)+4\triangledown_{\mu}\triangledown_{\nu}({\cal
F}_{2}(\Box_s)R^{\mu\nu})
\nonumber\\ &
+2(\Omega_{1\sigma}^{\;\sigma}+2\bar{\Omega}_{1})+2(\Omega_{2\sigma}^{\;\sigma}
+2\bar{\Omega}_{2})+2(\Omega_{3\sigma}^{\;\sigma}+2\bar{\Omega}_{3}
)-4\Delta_{2\sigma}^{\;\sigma}-8\Delta_{3\sigma}^{\;\sigma} \biggr)
\nonumber\\
= & -T\equiv
-g_{\al\bt}T^{\al\bt}\,.
\label{trace}
\end{align}
The Bianchi identity has been verified explicitly in Ref.~\cite{Biswas:2013cha}. Here we briefly sketch the Weyl part, since this will be the most important part of our discussion. 
To accomplish this,  note that the computations are simplified if one uses the following tricks by rewriting the equations of motion with one upper and one lower index, express Ricci tensor through the Einstein tensor (who's divergence is zero due to the Bianchi identities), and recalling the fact that the divergence of the Weyl tensor is the third rank Cotton tensor, which can be expressed  through the Schouten tensor:
\begin{equation*}
	\nabla^\gamma W_{\alpha\mu\nu\gamma}=-\nabla_\alpha S_{\mu\nu}+\nabla_\mu S_{\alpha\nu}\,,
\end{equation*}
where the Schouten tensor  in four dimensions is given by:
{
\begin{equation*}
	S_{\mu\nu}=\frac12\left(R_{\mu\nu}-\frac1{6}g_{\mu\nu}R\right)\,.
\end{equation*}
}
With this in mind the rest of the computations amount to careful accounting of the symmetry properties of the Weyl tensor (which are identical to those of the Riemann tensor). We should also observe  that the Bianchi identities should hold irrespectively of the precise form of functions $\mathcal{F}_i$, and independently for each and every coefficient $f_{i,n}$, because these are mere numerical coefficients, which are required to make the theory {\it ghost free}~\cite{Biswas:2013cha}\footnote{We have checked that the Bianchi identity holds true at each and every order of $\Box_s$.}
Technically, this means that we should not bother about the summation over $n$, but rather concentrating on the inner summation over $l$ in Eqs.~(\ref{details}-\ref{details-1}). Finally, the symmetry with respect to $\alpha\leftrightarrow\beta$ permutation in the equations of motion can be accounted by re-arranging the summation over $l$ in the inverse order from $n-1$ to 0. With all the above precautions in mind we can perform a direct substitution and check term by term that all the contributions vanish upon computing the divergence of the equations of motion. As stated above, it is not a surprise that the Bianchi identities hold. However, it is a very good check for the equations of motion, mostly for the mutual coefficients in front of the different terms, details can be found in Ref.~\cite{Biswas:2013cha}.

Let us note that in GR, we have a vacuum solution, around an {\it asymptotically Minkowski} background for a static case,
\begin{equation}
R=0\,,~~~~~~R_{\mu\nu}=0\,.
\end{equation}
In this case the energy momentum tensor vanishes in all the region except at $r=0$, where the source $m\delta^3(r)$ is localized. One of the properties of such a vacuum solution is the presence of $1/r$- static and spherically symmetric metric solution, similar to the Schwarzschild metric, given by:
{
\begin{equation}
ds^{2}=-b(r)dt^{2}+
b^{-1}(r) dr^{2} + 
r^{2} \left( d\theta^{2} + \sin^2(\theta) d\phi ^2\ \right)\; ,
\label{metric-1}
\end{equation}
}
where $b(r) =1-2Gm/r$ with the presence of a central singularity at $r=0$, and also the presence of an event horizon. As we have already discussed, for $r< 2Gm$, what dominates is the
$1/r$ part of $b(r)$, which dictates the rise in the gravitational potential all the way to $r=0$.
 Note that, although the vacuum solution permits $R=0,~R_{\mu\nu}=0$, the Weyl-tensor is non-vanishing in the case of a Schwarzschild metric, where $$W_{\mu\nu\lambda\sigma}W^{\mu\nu\lambda\sigma}\rightarrow \infty, $$ as $r\rightarrow 0$. Now in our case, indeed the full equations of motion are quite complicated, nevertheless, we might be able to test this hypothesis of setting $R=0,~R_{\mu\nu}=0$, and study whether the Schwarzschild metric, or $1/r$ type metric potential is a viable metric solution of our theory of gravity or not?

Let us then demand that the above action, Eq.~(\ref{action}), along with the equations of motion Eq.~(\ref{EOM}), permits a solution which is Schwarzschild metric with
$P_{\alpha\beta}=0$, and $R=0 $ and $R_{\mu\nu}=0$. In fact, in the region of non-locality where higher derivative terms in the action are dominant, it suffices to demand that $R={constant}$ and $R_{\mu\nu}={ constant}$. Let us now concentrate on the full equations of motion (\ref{EOM}) with the Weyl part of the full equations of motion:
 \begin{align}
 P^{\alpha\beta} = 0 =  P^{\alpha\beta}_3=& \frac{\alpha_c}{{ 8\pi G}} \biggl( -g^{\alpha\beta}W^{\mu\nu\lambda\sigma}{\cal
F}_{3}(\Box_s)W_{\mu\nu\lambda\sigma}+4W_{\;\mu\nu\sigma}^{\alpha}{\cal {\cal
F}}_{3}(\Box_s)W^{\beta\mu\nu\sigma}
\nonumber\\&
-4(R_{\mu\nu}+2\triangledown_{\mu}
\triangledown_{\nu})({\cal {\cal F}}_{3}(\Box_s)W^{\beta\mu\nu\alpha})
-2\Omega_{3}^{\alpha\beta}+g^{\alpha\beta}(\Omega_{3\gamma}^{\;\gamma}+\bar{
\Omega}_{3}) -8\Delta_{3}^{\alpha\beta} \biggr) \; .
\label{P3}
 \end{align}
Indeed, we would expect that  in order to fulfill the necessary condition (but not sufficient) for the Schwarzschild metric to be a solution of Eq.~(\ref{EOM}), 
we would have both the left and the right hand side of the above equation vanishes identically. The failure of this test will imply that the 
Schwarzschild metric {\it cannot} be the permissible solution of the equation of motion for Eq.~(\ref{EOM}).\\

There are couple of important observations to note, which we summarize below:

\begin{enumerate}
\item ${\cal F}_{i}(\Box_s)$ contain an infinite series of $\Box_s$.
\item The Bianchi identity holds for each and ever order in $\Box_s$, as we have already  discussed.
\item The right hand side of Eq.~(\ref{P3}) should vanish at each and every order in $\Box_s$. 
{This is due to the fact that when we compare the terms, assigned to coefficients $f_{i, n}$ (where the box operator has been applied $n$ times, i.e. $\Box_s^n$)
with terms where the box operator has been applied $n+1$ times ($\Box_s^{n+1}$, assigned to coefficient $f_{i, n+1}$), then the  $1/r^n$ dependence would at least be changed to $1/r^{n+2}$ in this process. Note that the box operator has roughly two covariant derivatives in $r$.}
Therefore, if we are not seeking for any miraculous cancellation, between different orders in $\Box_s$, it is paramount that each and every order in
$\Box_s$, the right hand side must vanish to yield the Schwarzschild-like metric solution. 

\item In fact, we could repeat the same argument for higher order singular metric ansatzs, such as $1/r^\alpha$, for $\alpha > 0$ at short distances, near the ultraviolet.\\
\end{enumerate}

In order to obtain some insight into this problem, let us first consider the right hand side of $P^{\alpha\beta}_3$ with one $\Box_s$ {\it only}, such that
$${\cal F}_{3}(\Box_s)=\left( f_{30}+f_{31}\Box_s \right).$$ Therefore, Eq.~(\ref{P3}) becomes
  \begin{align}
 P^{\alpha\beta}_{3} = & \frac{\alpha_c}{{8\pi G}} \biggl(-g^{\alpha\beta}W^{\mu\nu\lambda\sigma}(f_{30}+f_{31}\Box_s)W_{\mu\nu\lambda\sigma}
+4W_{\;\mu\nu\sigma}^{\alpha}(f_{30}+f_{31}\Box_s)W^{\beta\mu\nu\sigma}
\nonumber\\&
-4(R_{\mu\nu}+2\triangledown_{\mu}
\triangledown_{\nu})((f_{30}+f_{31}\Box_s)W^{\beta\mu\nu\alpha})
-2f_{31}\n^{\al}W^{\mu\nu\rho\ga}\n^{\bt}W_{\mu\nu\rho\ga}
\nonumber\\&
+g^{\alpha\beta}f_{31}(\n^{\al}W^{\mu\nu\rho\ga}\n^{\bt}W_{\mu\nu\rho\ga}+W^{
\mu\nu\rho\ga}\Box_s W_{\mu\nu\rho\ga})
\nonumber\\&
-8f_{31}(W^{\ga\nu}{}_{\rho\mu}\n^{\al}W_{\ga}{}^{\bt\rho\mu}-W_{\ga}{}^{
\bt\rho\mu}\n^{\al}W^{\ga\nu}{}_{\rho\mu})_{;\nu} \biggr) \: .
\label{P3-onebox}
 \end{align}
In the static limit,  after some computations, we can infer the following:

\begin{enumerate}

\item All the terms combining $f_{30}$ terms cancel each other from the above expression in Eq.~(\ref{P3-onebox}). This is indeed reminiscence, and agrees to 
the earlier computations performed in this regard in Ref.~\cite{Lu:2015cqa}, where the action corresponds to just the quadratic in curvature, but with local quadratic curvature action:
\begin{equation}\label{action-1}
S=\frac{1}{16\pi G}\int d^{4}x\sqrt{-g}\left({R}+\alpha_c [R^2+R^{\mu\nu}R_{\mu\nu}+ W^{\mu\nu\lambda\sigma}W_{\mu\nu\lambda\sigma}]\right)\,.
\end{equation}
Such an action indeed provides {\it singular} solutions with metric coefficients {$b(r) \sim 1/r$} for $r << r_{sch} $, as the leading order contribution, in spite of the fact that the 
above action has been shown to be renormalizable, but with an unstable vacuum, due to spin-2 ghost~\cite{-K.-S.}. The BGKM action indeed attempts to address
the ghost problem of quadratic curvature gravity.

\item The first non-trivial result comes from the fact that the {\it only} terms which {\it do not cancel}, and survive from the right hand side of Eq.~(\ref{P3-onebox}) are those proportional to $f_{31}$, and one can show explicitly that {they go as  $$1/r^8, $$} in the UV ( $r\ll 1/M_s$), for details see the Appendix. This means, that indeed $1/r$ as a metric 
solution does not pass through our test, since the right hand side of the above equation of motion is non-vanishing, but the left hand side ought to vanish in lieu of the vacuum condition, $P^{\alpha\beta}=0$.

\item In fact, we may be able to generalize our results to any orders in $\Box_s$
by noting that the higher orders beyond one box would contribute,
at least, two more covariant derivatives in $r$ 
in going from $\Box_s^n$ to $\Box_s^{n+1}$ terms 
(assuming that $\Box_s\thicksim \frac{1}{M_s^2} \partial _{r }^2$~~\footnote{
For the Schwarzschild metric this takes the form $
\Box_s = \frac{1}{M_s^2} g^{\nu\mu}\nabla_\nu\nabla_\mu = \frac{1}{M_s^2} \biggl[ \biggl(1-\frac{2m}{r}\biggr)\partial _{r }^2 - 2\biggl( -\dfrac{m}{r^2} + \biggl(1-\frac{2m}{r}\biggr)\dfrac{1}{r}\biggr) \partial _{r } \biggr]
$.}). This means that the full computation for the right hand side of Eq.~(\ref{P3-onebox}) would yield:
\begin{equation}
P^{\alpha\beta}_{3}\backsim g^{\alpha\beta}\biggl( f_{31} {\cal O}(\frac{1}{r^8}) + f_{32} {\cal O}(\frac{1}{r^{10}})+...+ f_{3n}{\cal O}(\frac{1}{r^{6+2n}})+...\biggr)\,,
\end{equation}
($g^{\alpha\beta}$ is defined from the metric \eqref{metric-1}, see the exact definition of $P^{\alpha\beta}_{3}$ in the appendix) which would require too much fine tuning to cancel each and every term, while keeping in mind that $f_{3n}$ are mere constant coefficients. Barring such unjustified cancellation, it is fair to say that indeed $1/r$ for $r \ll 1/M_s$ as a metric potential for the BGKM gravity is not a valid solution, if ${\cal F}_3$ has a nontrivial dependence on $\Box_s$.\\

\end{enumerate}

Similar conclusions have already been drawn in Ref.~\cite{Buoninfante:2018xiw}, with a complementary arguments. In Ref.~\cite{Buoninfante:2018xiw}, the argument was based on taking a smooth limit from non-linear solution of Eq.~(\ref{action}) to the linear solution. For any physical solution to be valid, the non-linear solution must pave to the linear solution smoothly.

At the linear level (where the metric potential is bounded below 1),  it was shown that the Weyl term vanishes quadratically in $r$~\cite{Buoninfante:2018xiw}, for a non-singular metric solution given by a metric potential Eq.~(\ref{erf}). Therefore, at the full non-linear level $1/r$-type metric potential cannot be promoted as a full solution for the non-linear
equations of motion for the BGKM action, since there is no way it can be made to vanish quadratically at the linear level. Similar conclusions can be made for any metric potential which goes as $1/r^{\alpha}$ for $\alpha > 1$.

Indeed, this intriguing and potentially very powerful conclusion leads to the fact that the BGKM action with quadratic curvature, infinite covariant derivative gravitational action not only ameliorates the curvature singularity at $r=0$, but also gives rise to a metric potential which is bounded below one in the entire spacetime regime. The notion about the physical mechanism which avoids forming a trapped surface, also yields a static metric solution of gravity, which has no horizon, see \cite{Frolov-book}.
The only viable solution of Eq.~(\ref{action}) remains that of the linear solution, around the Minkowski background, already described by Eq.~(\ref{erf}). Indeed, this last step has to be shown more rigorously, which we leave for future investigation.

Another important conclusion arises due to the non-local interactions in the gravitational sector, which yields a non-vacuum solution, such that $R\neq 0$ and $R_{\mu\nu}\neq 0$, within length scale 
$\sim 2/M_{s}$~\cite{Buoninfante:2018xiw}. This is due to the fact that the BGKM gravity smears out the Dirac-delta source, and therefore a vacuum solution does not exist any more like in the case of the Einstein's gravity, or  any $f(R)$ gravity, or even in the context of  local quadratic curvature gravity, see Eq.~(\ref{action-1}).

\section{conclusion}

The conclusion of this paper is very powerful. We have argued that Schwarzschild metric or $1/r$-type metric potentials cannot be the solution of the full BGKM action given by Eq.~(\ref{action}), and the full non-linear equations of motion~Eq.~\ref{EOM}). By $1/r$-type metric potential, we mean the non-linear part of the Schwarzschild metric, for $ r< 2Gm$, where
$m$ is the Dirac delta source.  The presence or absence of singularity is judged by the Weyl contribution. In the pure Einstein-Hilbert action, indeed the Weyl term in the Schwarzschild metric is non-vanishing, and contributes towards the Kretschmann singularity at $r=0$. In the case of infinite derivatives in $4$ dimensions, we have shown here that this is not the case, and the infinite derivative Weyl contribution contradicts with $1/r$ being the metric solution for a vacuum configuration, for which the energy momentum tensor vanishes, for a static and spherically symmetric solution for the BGKM action. By itself the result does not prove or disprove a non-singular metric potential, but it provides a strong hint that the full equations of motion cannot support Schwarzschild type of $1/r$ type metric potential. We have also argued that on a similar basis even $1/r^{\alpha}$ for $\alpha >0$ will not serve as a full solution to the BGKM gravity. It would be very interesting to explore that  if the BGKM gravity may allow other static/non-static singular metric solutions or not~\cite{Buoninfante:2018xif}.


\section{Appendix}

\subsection{Explicit non-vanishing contributions from the Weyl term}

Here we show the relevant terms, present in Eq. (\ref{P3-onebox}), assuming {$b(r) = 1- 2Gm/r$} in the metric (\ref{metric-1}). The explicit enumeration of each term is important to understand how the coefficient $f_{30}$ and $f_{31}$ appear and how they might cancel. Let us define 
$$P_3^{\alpha \beta }= \frac{\alpha_c}{{ 8\pi G}} \sum_{i=1}^{6} F_i^{\alpha \beta }$$

\begin{enumerate}
\item For the first term, $F_1^{\alpha \beta }=-g^{\alpha\beta}W^{\mu\nu\lambda\sigma}(f_{30}+f_{31}\Box_s)W_{\mu\nu\lambda\sigma}$, the calculation yields
\begin{equation}
F_1^{\alpha \beta }= g^{\alpha \beta}\frac{48G^2}{r^8} \frac{m^2}{M_{s}^2}\left(
\begin{array}{cccc}
- \frac{ \left(f_{30} M_{s}^2 r^3 - 6f_{31} G m  \right)}{r} & 0 & 0 & 0 \\
 0 & -\frac{ \left( f_{30} M_{s}^2 r^3 - 6f_{31} G m \right)}{r} & 0 & 0 \\
 0 & 0 & -\frac{ \left( f_{30} M_{s}^2 r^3 - 6f_{31} G m \right)}{ r} & 0 \\
 0 & 0 & 0 & -\frac{ \left( f_{30} M_{s}^2 r^3 - 6f_{31} G m \right) }{ r} \\
\end{array}
\right)\: .
\end{equation}

\item The second term,
$F_2^{\alpha \beta }= +4 W^{\alpha }{}_{\mu \nu \sigma } \left(f_{30} + f_{31} \Box_s \right) W^{\beta \mu \nu \sigma }$, is given by
\begin{equation}
F_2^{\alpha \beta }=- g^{\alpha \beta}\frac{48G^2}{r^8} \frac{m^2}{M_{s}^2}\left(
\begin{array}{cccc}
- \frac{ \left(f_{30} M_{s}^2 r^3 - 6f_{31} G m  \right)}{r} & 0 & 0 & 0 \\
 0 & -\frac{ \left( f_{30} M_{s}^2 r^3 - 6f_{31} G m \right)}{r} & 0 & 0 \\
 0 & 0 & -\frac{ \left( f_{30} M_{s}^2 r^3 - 6f_{31} G m \right)}{ r} & 0 \\
 0 & 0 & 0 & -\frac{ \left( f_{30} M_{s}^2 r^3 - 6f_{31} G m \right) }{ r} \\
\end{array}
\right)\: .
\end{equation}
We can verify, at this point, that the first two terms cancel each other. 

\item The third term,
$F_3^{\alpha \beta }= -4\left(2 R_{\mu \nu }+\nabla _{\mu }\left.\nabla _{\nu }\right) \left( f_{30} + f_{31} \Box_s \right)W^{\beta \mu \nu \alpha }\right.$, is given by
\begin{equation}
F_3^{\alpha \beta }= g^{\alpha \beta} \frac{288G^2}{r^8} \frac{m^2}{M_{s}^2}f_{31} \left(
\begin{array}{cccc}
 -\frac{ (5r-11Gm)}{r} & 0 & 0 & 0 \\
 0 & -\frac{(r-3Gm)}{r} & 0 & 0 \\
 0 & 0 & \frac{(3r-7Gm)}{r} & 0 \\
 0 & 0 & 0 & \frac{(3r-7Gm)}{r} \\
\end{array}
\right)\: ,
\end{equation}
which only depends on the $f_{31}$ coefficient.

\item The fourth term, $F_4^{\alpha \beta }=  -2f_{31}\nabla ^{\alpha }W^{\lambda }{}_{\mu \nu \sigma }\nabla ^{\beta }W_{\lambda }^{\:\: \mu \nu \sigma } $, is given by
\begin{equation}
F_4^{\alpha \beta }= g^{\alpha \beta} \frac{288G^2}{r^8} \frac{m^2}{M_{s}^2}f_{31} \left(
\begin{array}{cccc}
 0 & 0 & 0 & 0 \\
 0 & -\frac{ 3(r-2Gm) }{r} & 0 & 0 \\
 0 & 0 & -\frac{ (r-2Gm) }{r} & 0 \\
 0 & 0 & 0 & -\frac{ (r-2Gm) }{r } \\
\end{array}
\right)\: .  
\end{equation}

\item The fifth term, $F_5^{\alpha \beta }=  +g^{\alpha\beta}f_{31}(\n^{\al}W^{\mu\nu\rho\ga}\n^{\bt}W_{\mu\nu\rho\ga}+W^{\mu\nu\rho\ga}\Box_s W_{\mu\nu\rho\ga}) $, is given by
\begin{equation}
F_5^{\alpha \beta }= g^{\alpha \beta} \frac{144G^2}{r^8} \frac{m^2}{M_{s}^2}f_{31}  \left(
\begin{array}{cccc}
 \frac{ (5 r-12Gm) }{r} & 0 & 0 & 0 \\
 0 & \frac{ (5 r-12Gm) }{r} & 0 & 0 \\
 0 & 0 & \frac{ (5 r-12Gm) }{r} & 0 \\
 0 & 0 & 0 & \frac{ (5 r-12Gm) }{r } \\
\end{array}
\right)
\: .
\end{equation}

\item The sixth term, $F_6^{\alpha \beta }= -8f_{31}(W^{\ga\nu}{}_{\rho\mu}\n^{\al}W_{\ga}{}^{\bt\rho\mu}-W_{\ga}{}^{\bt\rho\mu}\n^{\al}W^{\ga\nu}{}_{\rho\mu})_{;\nu}  $, is given by
\begin{equation}
F_6^{\alpha \beta }= g^{\alpha\beta} \frac{576 G^2}{r^8} \frac{m^2}{M_{s}^2}f_{31} 
\left(
\begin{array}{cccc}
 0 & 0 & 0 & 0 \\
 0 & -\frac{ (r-2Gm) }{r} & 0 & 0 \\
 0 & 0 & \frac{ (3 r-7Gm) }{r} & 0 \\
 0 & 0 & 0 & \frac{ (3 r-7Gm) }{r } \\
\end{array}
\right)
\: .
\end{equation}
\end{enumerate}

Having computed each term of  $P^{\alpha\beta}_{3}$, we can conclude that the stress energy momentum tensor dependence on 
the $f_{30}$ coefficient is vanishing, and {\it only} the one box, $\Box_s$, contributions survive. 
Finally, we have the non vanishing contribution,
\begin{equation}
P_3^{\alpha \beta }= g^{\alpha\beta} \frac{144 { G} }{8\pi r^8} \frac{m^2}{M_{s}^4}f_{31} 
\left(
\begin{array}{cccc}
 - \frac{5 (r-2Gm) }{r}& 0 & 0 & 0 \\
 0 & -\frac{7 (r-2Gm) }{r} & 0 & 0 \\
 0 & 0 & \frac{ (21r-50Gm) }{r} & 0 \\
 0 & 0 & 0 & \frac{ (21 r-50Gm) }{r } \\
\end{array}
\right)
\: .
\label{P-1box}
\end{equation}

{\bf Second order in $\Box_s$ contributions}:
In order to strengthen our arguments, 
we present below the additional contribution for the second order in box, i.e., $\Box_s^2$: 
\begin{equation}
	P^{\alpha\beta}_3(\Box_s^2)= - g^{\alpha\beta} \frac{576 { G} }{8\pi r^{10}} \frac{m^2}{M_{s}^6}f_{32}
\left(
\begin{array}{cccc}
 a_{00}& 0 & 0 & 0 \\
 0 & a_{11} & 0 & 0 \\
 0 & 0 & a_{22} & 0 \\
 0 & 0 & 0 & a_{33} \\
\end{array}
\right)
\: . 
\label{P-2box}
\end{equation}
with the dimensionless matrix elements, defined as
\begin{align}
a_{00} =&\frac{ (939 G^2 m^2-744 G m r+140 r^2) }{r^2}\;, \nonumber \\
a_{11} =& \frac{ (195 G^2 m^2 - 132 G m r + 20 r^2) }{r^2}\;, \nonumber \\
a_{22} =& -\frac{ (789 G^2 m^2-534 G m r+80 r^2) }{r^2}\;, \nonumber \\
a_{33} =& - \frac{ (789 G^2 m^2-534 G m r+80 r^2) }{r^2}\;. \nonumber
\end{align}
Let us now consider, for example, the $P^{22}_{3}$ element at $\Box_s^2$, namely:
\begin{equation}
P^{22}_{3} = \frac{\alpha_c}{{ 8\pi G}} \biggl[f_{31}\left(\frac{3024 G^2 m^2}{r^{10} M_s^2}-\frac{7200 G^3 m^3 }{r^{11} M_s^2}\right)+
f_{32}\left( \frac{46080 G^2 m^2 }{r^{12} M_s^4}-\frac{307584 G^3 m^3 }{r^{13} M_s^4}+\frac{454464 G^4 m^4}{r^{14} M_s^4}\right) + \cdots\biggr]\: .
\end{equation}
Demanding that $P^{22}_{3}=0$, implies that $f_{31}=f_{32}=0$. We can now ask what would happen for higher orders in $\Box_s$. Since  $\Box_s\thicksim \frac{1}{M_s^2} \partial _{r }^2$, we have at the lowest third order contribution in box, in powers of $r$, is proportional to
\begin{equation}
f_{33} \frac{G^2 m^2 }{r^{14} M_s^6}\,.
\end{equation}
Therefore, since we already concluded that $f_{31}=f_{32}=0$, we now have to demand that the contribution of $f_{33} \frac{G^2 m^2 }{r^{14} M_s^6}$ vanishes identically. The lowest fourth order contribution, in powers of $r$, will go as
\begin{equation}
f_{34} \frac{G^2 m^2 }{r^{16} M_s^8}\,,
\end{equation}
we are left with the option that $f_{33}=f_{34}=0$. Obviously, we do not claim that this is a rigorous mathematical demonstration, however we can hint, by dimensional analysis, that the lowest $n^{nth}$ order contribution will be always proportional to
\begin{equation}
f_{3n} \frac{G^2 m^2 }{r^{8+2n} M_s^{2n}}\,.
\end{equation} 
Indeed,  the aforementioned analysis will make  it hard to make all the  expansion coefficient set to vanish, i.e. $f_{3n}=0$.

\acknowledgments 

We would like to thank Luca Buoninfante, Tirthabir Biswas, Valeri Frolov, Tomi Koivisto, and Robert Brandenberger for numerous discussions at various stages of this project.
AK and JM are supported by the grant UID/MAT/00212/2013 and COST Action CA15117 (CANTATA). AK is supported by FCT Portugal investigator project IF/01607/2015 and FCT Portugal fellowship SFRH/BPD/105212/2014. AM's research is financially supported by Netherlands Organisation for Scientific Research (NWO) grant number 680-91-119.



\end{document}